\documentclass[twocolumn,showpacs,showkeys,superscriptaddress]{revtex4}

\usepackage{amsmath}
\usepackage{bbold}
\usepackage{amsfonts}
\usepackage{amssymb}
\usepackage{pbsi}
\usepackage[T1]{fontenc}
\usepackage{xcolor}
\usepackage{graphicx}
\usepackage{subfig}
\usepackage{float}
\usepackage{color}
\usepackage{hyperref}
\hypersetup{colorlinks = true,
	linkcolor = red,
	anchorcolor = blue,
	citecolor = blue,
	filecolor = blue,
	urlcolor = blue}

\begin{document}

\title{Time-domain supersymmetry for massless scalar and electromagnetic  fields in anisotropic cosmologies}

\author{Felipe A. Asenjo}
\email{felipe.asenjo@uai.cl}
\affiliation{Facultad de Ingenier\'ia y Ciencias,
Universidad Adolfo Ib\'a\~nez, Santiago 7491169, Chile.}
\author{Sergio A. Hojman}
\email{sergio.hojman@uai.cl}
\affiliation{Departamento de Ciencias, Facultad de Artes Liberales,
Universidad Adolfo Ib\'a\~nez, Santiago 7491169, Chile.}
\affiliation{Departamento de F\'{\i}sica, Facultad de Ciencias, Universidad de Chile,
Santiago 7800003, Chile.}

\date{\today}

\begin{abstract}
It is shown that any cosmological anisotropic model produces supersymmetric theories for both massless scalar and electromagnetic fields. This  supersymmetric theory is the time-domain analogue  of a supersymmetric quantum mechanical theory. In this case, the variations of the anisotropic scale factors of the Universe are responsible for triggering the supersymmetry. For scalar fields, the superpartner fields evolve in two different cosmological scenarios (Universes). On the other hand, for propagating  electromagnetic fields, supersymmetry is manifested through its polarization degrees of freedom in  one Universe. In this case, polarization degrees of freedom of  electromagnetic waves, which are orthogonal to its propagation direction, become superpartners from each other. This behavior can be measured, for example, through the rotation of the plane of polarization of cosmological light.
\end{abstract}



\maketitle

\section{Introduction}

Supersymmetry is one of the cornerstone concepts in all physics \cite{Freund,aiit,qcosmo1}. Recently, the concept of supersymmetry in time-domain (T-SUSY) appeared when describing light propagation on anisotropic cosmological backgrounds \cite{ah1} and in the eikonal limit in time dependent refractive index media \cite{carlos}. This kind 
of supersymmetry occurs for a system evolving in time, instead of space, having all the usual supersymmetric features of a quantum mechanical system. Therefore, this idea of T-SUSY
is applicable to any system of dynamical fields that are driven by some time-dependent potential. This was later extended to the relativistic regime of quantum mechanics \cite{inaoe}.

In this context, it is reasonable to ask ourselves about this supersymmetric features for fields evolving on a cosmological scenario. The aim of this work is to show that a T-SUSY theory can be formulated for massless scalar and electromagnetic fields propagating on
a general background anisotropic cosmological model. For both these fields, we show that supersymmetry occurs always for anisotropic Universes, where the superpotential of the supersymmetric theory depends on the scale factors of the metric, which are time dependent.  

In Secs. II and III we show that the T-SUSY behavior of massless scalar fields is different from the one exhibited by electromagnetic fields. While scalar fields require superpartners that evolve in two different anisotropic Universes, the polarizations of an electromagnetic field (light) are the superpartners of each other 
of this theory, evolving in just one anisotropic cosmology. This is analogue to what occurs in anisotropic media \cite{asenjomedia}.
In Sec. IV we discuss the implications. 

In order to start with the model, let us consider a general anisotropic
cosmology. Using the four-coordinates $x^\mu=(t,x,y,z)$ to describe this Universe, we have that the spacetime interval is $ds^{2}=g_{\mu\nu}dx^{\mu}dx^{\nu}=-dt^2+a(t)^2 dx^{2}+b(t)^2 dy^2+ c(t)^2 dz^2$, with the metric \cite{ryanshep}
\begin{eqnarray}
g_{\mu\nu}=\left(-1,a^2,b^2,c^2\right)\, ,
\label{anisotrometric}
\end{eqnarray}
where, in general, it is possible to have $a\neq b\neq c\neq a$, being all of them the different scale factors for each anisotropic direction of the Universe (here $\mu,\nu=0,1,2,3$). Different  exact solutions have been found for these kind of cosmologies \cite{gron}. Therefore, any of these solutions can be used as a cosmological background for the supersymmetric dynamics of  massless scalar and electromagnetic  fields that are studied below. The case when $a= b=c$ is the usual flat spatial Friedman-Lema\^itre-Robertson-Walker (FLRW) cosmology.

\section{Scalar fields}

A massless scalar field (let us call it $\phi_-$ for reasons that will become apparent later) in this anisotropic cosmology will be modelled by the Klein-Gordon equation 
\begin{equation}
    \Box\phi_-=\frac{1}{\sqrt{-g}}\partial_\mu\left(\sqrt{-g}g^{\mu\nu}\partial_\nu \phi_-\right)=0\, ,
\end{equation}
where we do not assume any modification to the minimal coupling prescription  to get a conformally invariant equation \cite{wald}. Considering that the field has a spatial dependence along an arbitrary direction, say the $z$-direction, then, using metric  \eqref{anisotrometric},
the previous equation simplifies to
\begin{equation}
    \frac{c}{ab}\frac{\partial}{\partial t}\left(a b c\frac{\partial\phi_-}{\partial t}\right)- \frac{\partial^2\phi_-}{\partial z^2}=0\, .
    \label{eq1kG}
\end{equation}
We can define the conformal time
$\tau=\int {dt}/{c}$, and assume a  spatial behavior in the form $\partial_z^2\phi_-=  -k^2\phi_-$, with an arbitrary constant $k$. Also, performing the change of variable
\begin{equation}
    \varphi_-=\sqrt{ab}\, \phi_-\, ,
\end{equation}
then Eq.~\eqref{eq1kG} is now written as
\begin{equation}
    \ddot\varphi_- +\left(-\dot W_s-W_s^2+k^2 \right)\varphi_-=0\, ,
    \label{susy1}
\end{equation}
where the overdot stands for $\tau$-derivatives, and 
\begin{equation}
    W_s=\frac{\dot a}{2a}+\frac{\dot b}{2b}\, .
\end{equation}
Eq.~\eqref{susy1} is the paradigmatic starting point of our analysis. The T-SUSY theory can be constructed by defining the two supersymmetry operators
\begin{equation}
    Q^s_\pm=\frac{d}{d \tau}\pm W_s\, .
\end{equation}
In this way, 
Eq.~\eqref{susy1}
is put in the form
\begin{eqnarray}
   H^s_- \varphi_-\equiv Q^s_+Q^s_-\varphi_-=-k^2\varphi_-\, ,
\end{eqnarray}
with the operator $H_-^s=Q^s_+Q^s_-$, and where the term $\dot W_s+W_s^2$ plays the role of the T-SUSY superpotential of the dynamics.

In order to introduce the supersymmetry in time-domain, the previous equation allows us to infer the existence of a supersymmetric massless scalar field partner
$\varphi_+$. Therefore, both fields $\varphi_\pm$ must satisfy the equations
\begin{equation}
    Q^s_\pm \varphi_\pm=i k\varphi_\mp\, ,
    \label{superKGpart}
\end{equation}
such that the superpartner $\varphi_+$ fulfills
\begin{eqnarray}
   H^s_+ \varphi_+\equiv Q^s_-Q^s_+\varphi_+=-k^2\varphi_+\, ,
\end{eqnarray}
with the operator $H_+^s=Q^s_-Q^s_+$.
This last equation translates to
\begin{equation}
    \ddot\varphi_+ +\left(\dot W_s-W_s^2+k^2 \right)\varphi_+=0\, ,
    \label{susy2}
\end{equation}
where the the partner superpotential $-\dot W_s+W_s^2$ emerges. 
In this way, by defining 
the field
\begin{eqnarray}
    \phi_+=\sqrt{ab}\, \varphi_+\, ,
\end{eqnarray}
we get, from \eqref{susy2},  that $\phi_+$ must satisfy the following scalar massless equation
\begin{equation}
    {abc}\frac{\partial}{\partial t}\left(\frac{c}{ab}\frac{\partial\phi_+}{\partial t}\right)- \frac{\partial^2\phi_+}{\partial z^2}=0\, ,
    \label{eq2kG}
\end{equation}
in time $t$. This implies that $\phi_+$ is a massless scalar field with a dynamic that is developing in an anisotropic Universe with metric
\begin{eqnarray}
g_{\mu\nu}=\left(-1,\frac{1}{a^2},\frac{1}{b^2},c^2\right)\, .
\label{anisotrometric2}
\end{eqnarray}

In this sense, field $\varphi_+$ ($\phi_+$) is the superpartner of $\varphi_-$ ($\phi_-$), which is explicitly displayed by Eqs.~\eqref{superKGpart}.
They do not evolve in the same Universe, but rather in different cosmological scenarios. In this way, T-SUSY cannot be observed for massless
scalar fields in our (solely) Universe. 

Furthermore, the above results imply that the supersymmetric behavior for massless scalar field always occur in  anisotropic spacetimes, even for one isotropic Universe. 

By studying the massless field $\phi_-$ for the isotropic cosmological Universe described by the FLRW metric, we inevitable arrive to that the superpartner field $\phi_+$ must evolve in an anisotropic cosmology with metric $(-1,a^{-2},a^{-2},a^{2})$. Thus, the cosmological T-SUSY behavior for massless scalar fields
requires that Universes be anisotropic.

\section{Electromagnetic fields}

Vectorial massless field have a very different behavior to the previous case. Supersymmetry in space-domain has been 
already proposed for light in the eikonal limit \cite{kwolf,MatthiasHeinrich,miri1,miri2,Hokmabadi1,machollor,HeinrichHeinrich,miri3}. On the other hand, the T-SUSY behavior of electromagnetic field in anisotropic cosmolgies was first suggested in the Appendix of Ref.~\cite{ah1}. In here we extend this idea and analysis in a thoroughly manner.

Maxwell equations in curved spacetimes read $\nabla_\alpha F^{\alpha\beta}=0$, which can be written as
\begin{eqnarray}
    \frac{1}{\sqrt{-g}}\partial_\alpha\left[\sqrt{-g} g^{\alpha\mu}g^{\beta\nu}(\partial_\mu A_\nu-\partial_\nu A_\mu) \right]=0\, ,
    \label{Maxwell1}
\end{eqnarray}
where $\nabla_\mu$ is a covariant derivative,
$F_{\mu\nu}=\partial_\mu A_\nu-\partial_\nu A_\mu$, is the electromagnetic tensor, and $A_\mu$ is the electromagnetic potential vector.
Now, let us consider a transversal electromagnetic field, such that its polarization is perpendicular to its spatial propagation. Thus, without loss of generality, let us assume that $A_x=A_x(t,z)$ and $A_y=A_y(t,z)$
are the only non-vanishing components of the electromagnetic potential vector. This choice also fulfill the Lorentz gauge, $\nabla_\mu A^\mu=0$.
Therefore, using the anisotropic cosmology metric \eqref{anisotrometric} in Eq.~\eqref{Maxwell1}, we get \cite{ah1}
\begin{eqnarray}
    \frac{a c}{b}\frac{\partial}{\partial t}\left( \frac{b c}{a}\frac{\partial A_x}{\partial t}\right)-\frac{\partial^2 A_x}{\partial z^2}
&=&0\, ,\nonumber\\
\frac{b c}{a}\frac{\partial}{\partial t}\left( \frac{a c}{b}\frac{\partial A_y}{\partial t}\right)-\frac{\partial^2 A_y}{\partial z^2}
&=&0\, .
\end{eqnarray}

Now, let us allow  to change the name of variables to $A_x\equiv A_+$ and 
$A_y\equiv A_-$. Also, let us consider  a  spatial behavior in the form $\partial_z^2A_{x,y}=  -k^2 A_{x,y}$, with an arbitrary constant $k$.
In this case, both previous equations acquire the simple form
\begin{eqnarray}
    \xi^{\mp 1}\frac{\partial}{\partial \tau}\left(\xi^{\pm 1} \frac{\partial A_\pm}{\partial \tau}\right)+k^2 A_\pm=0\, ,
    \label{Maxw2}
\end{eqnarray}
where we have used again the conformal time
$\tau=\int {dt}/{c}$, and the variable
\begin{eqnarray}
    \xi=\frac{b}{a}\, .
\end{eqnarray}
Note that, in general, $\xi$ only varies in an 
anisotropic cosmology in a transversal plane to the propagation of the electromagnetic field. 

Now, by defining 
\begin{eqnarray}
   {\cal A}_{\pm}=\xi^{\pm 1/2}A_\pm\, ,
\end{eqnarray}
we found, using Eq.~\eqref{Maxw2}, the T-SUSY set of equations
\begin{eqnarray}
    \ddot {\cal A}_\pm+ \left(k^2- W_e^2\pm \dot W_e\right) {\cal A}_\pm=0\, ,
    \label{eqsusymaxwel1}
\end{eqnarray}
where the overdot is the conformal time derivative, and 
\begin{eqnarray}
    W_e=-\frac{\dot\xi}{2\xi}=\frac{\dot a}{2 a}-\frac{\dot b}{2b}\, .
    \label{weee}
\end{eqnarray}
Clearly, Eq.~\eqref{eqsusymaxwel1} can be put in the form
\begin{eqnarray}
    H^e_\pm {\cal A}_\pm\equiv Q^e_\mp Q^e_\pm {\cal A}_\pm=-k^2 {\cal A}_\pm\, ,
\end{eqnarray}
with the operators
\begin{equation}
    Q^e_\pm=\frac{d}{d \tau}\pm W_e\, ,
\end{equation}
and $H_\pm^e=Q_\mp^e Q_\pm^e$, and their respective superpotentials. 
Thus, it is straightforward to construct the set of supersymmetric equations
\begin{equation}
    Q^e_\pm {\cal A}_\pm=i k {\cal A}_\mp\, .
    \label{Maxfirstorderlight}
\end{equation}
Thus, both transversal polarizations  of an electromagnetic wave become superpartners of each other, as it propagates, evolving in a cosmological time. This occurs for any kind of cosmological anisotropy, as $\dot W_e\neq 0$ 
for $a\neq b$. 

Differently to what occur for scalar massless fields, the T-SUSY behavior of  light occurs in a single cosmological Universe. Any anisotropy of the cosmological expansion of the Universe, as small as it may be, will trigger this supersymmetric behavior on the polarization of light. This is a consequence of the existence of solutions which exhibit non-null propagating behavior of light in anisotropic cosmologies (or in any media) \cite{ah1}.
Therefore, any  measurement of supersymmetry in cosmological light can bring insights about the anisotropy
of our Universe.

\section{Discussion}

As well as for massless scalar fields as for electromagnetic fields, the above description of supersymmetric behavior is analogue to those for  quantum mechanical systems.
Thereby, 
  the framework for a  supersymmetric theory in quantum mechanics can be  utilized straightforwardly. For  both (scalar and electromagnetic) massless fields, we can  define  the super-Hamiltonian matrix operator ${\bf H}^j$ (for $j=s$ or $e$), and the super-charge matrix operators ${\bf Q}^j$  as \cite{coop2}
\begin{equation}
{\bf H}^j=\left(\begin{array}{cc}
   H_+^j & 0 \\
    0 & H_-^j
\end{array}\right),  \,  
 {\bf Q}^j=\left(\begin{array}{cc}
       0 & 0 \\
        Q_+^j & 0
\end{array}\right), \, 
{\bf Q}^{j \dag}=\left(\begin{array}{cc}
       0 & Q_-^j \\
        0 & 0
    \end{array}\right).
\end{equation}
These operators have the closed algebra ${\bf H}=\{{\bf Q},{\bf Q}^\dag\}$, $[{\bf H},{\bf Q}]=0=[{\bf H},{\bf Q}^\dag]$, $\{{\bf Q},{\bf Q}\}=0=\{{\bf Q}^\dag,{\bf Q}^\dag\}$ for the bosonic ${\bf H}$, and fermionic ${\bf Q}$ and ${\bf Q}^\dag$ operators.
In the case $k=0$, for instance, when both fields evolve on  cosmological time, we  can infer that the (scalar and electromagnetic) superpartners have  related eigenfunctions and eigenvalues \cite{coop2}.

A supersymmetric quantum theory, and its consequences, are standard notions. 
However, the main idea of this work is to show how anisotropies induce supersymmetric behavior on cosmological polarized electromagnetic waves. This supersymmetric property could be detected in our Universe using light propagation. 
For this supersymmetric solution, the electric and magnetic field components are $E_\pm=\partial_0 A_\pm=ik c\xi^{\mp 1/2}{\cal A}_\mp$, and $B_\pm=\mp \partial_z A_\mp=\mp ik \xi^{\mp 1/2}{\cal A}_\mp$, respectively. Their properties can be measured through the Stokes parameters \cite{hecht}.
For instance, in the case of a small anisotropy, such that $\xi\approx 1-2\epsilon$, where $\epsilon$ is the measure of the transversal anisotropy to light propagation (with
$\ddot\epsilon\neq 0$), then polarization solutions of Eqs.~\eqref{Maxfirstorderlight}
are ${\cal A}_\pm\approx \exp(i k \tau\mp S)$, where 
\begin{eqnarray}
    S= e^{-2ik\tau} \int  \dot\epsilon\,  e^{2 ik\tau}d \tau\, .
    \label{soluperturb}
\end{eqnarray}
This solution represents a quasi-monochromatic electromagnetic wave. Therefore, T-SUSY
produces 
a shift on the phase of the polarization of the  wave of the order
$2\cos(2k \tau)\int \dot\epsilon \sin(2k\tau)d\tau -2\sin(2k \tau)\int \dot\epsilon \cos(2k\tau)d\tau$. This
induces a rotation
of the plane of polarization of the electromagnetic wave at it propagates.  

These, among other features that could be obtained for polarizations satisfying \eqref{Maxfirstorderlight}, 
can be used to obtain, from measurements, any trace of this cosmological supersymmetric behavior of light.

\begin{acknowledgements}
FAA thanks to FONDECYT grant No. 1230094 that supported this work.
 \end{acknowledgements}

\end{document}